\input phyzzx
\newcount\mongocount
\mongocount=1
\def\Figure#1#2#3{
% \boxit{
      \vbox to #3in{\hsize=#2in
        \vfil
%\special{ps::[begin]
%          save 10 dict begin /Figure exch def
%          currentpoint translate
%          /showpage {} def
%        }
%        \special{ps: plotfile #1}
         \includegraphics{#1}
%        \special{ps::[end]
%        clear Figure end restore
%        }
    }
% }
}
\def\figcap#1#2{
\vtop{\tenpoint\singlespace
\hsize=#1in\smallskip\noindent Figure\ \ \the\mongocount.\ \  #2
\global\advance\mongocount by 1\bigskip}}
\def\mongofigure#1#2#3#4#5{\centerline{\Figure{#1}{#2}{#3}
\figcap{#4}{#5}}}

\hoffset=0.375in
\overfullrule=0pt

\def\CMD{{\rm CMD}}
\def\cmd{{\rm cmd}}
\def\exp{{\rm exp}}

\def\obs{{\rm obs}}

\def\pc{{\rm pc}}
\def\kpc{{\rm kpc}}
\def\kms{{\rm km}\,{\rm s}^{-1}}

\twelvepoint
\font\bigfont=cmr17
\centerline{\bigfont The Spheroid Luminosity and Mass Functions From}
\smallskip
\centerline{\bigfont Hubble Space Telescope Star Counts}
\smallskip
\centerline{{\bf Andrew Gould}\foot{Alfred P.\ Sloan Foundation Fellow}}
\smallskip
\centerline{Dept of Astronomy, Ohio State University, Columbus, OH 43210}
\smallskip
\smallskip 
\centerline{\bf Chris Flynn}
\smallskip
\centerline{Tuorla Observatory, Piikki\"{o}, FIN-21500, Finland}
\smallskip
\centerline{\bf John N.\ Bahcall}
\smallskip
\centerline{Institute For Advanced Study, Princeton, NJ}
\smallskip
\centerline{e-mail: gould@payne.mps.ohio-state.edu, 
cflynn@astro.utu.fi,jnb@ias.edu}

\centerline{\bf Abstract}
\singlespace

We analyze 166 spheroid subdwarfs $(6.5<M_V<14.5)$ found in 53 fields observed 
with the Wide Field Camera on the {\it Hubble Space Telescope}.  The 
fields cover 221 arcmin$^2$ over a wide range of directions.  
The spheroid 
luminosity function (LF) is inconsistent at about the $3\,\sigma$ level
with the local spheroid LF of Dahn et al.\  even when the normalization of 
the latter is corrected to take account of the latest data on spheroid 
kinematics.  The difference may reflect systematic errors in one of the two
studies or features of the spheroid spatial distribution that are not
included in the simplest models.
The mass function, which shows no obvious structure, can be represented by
a power law,
$d N/d\ln M\propto M^{\alpha}$, with $\alpha=0.25\pm 0.32$ over the mass 
range $0.71\,M_\odot> M >0.09\,M_\odot$.  The spheroid therefore does not
contribute significantly to microlensing unless the mass function changes
slope dramatically in the substellar range.  The total local mass density of
spheroid stars (including remnants and unseen binary companions) is 
$\rho\sim 6.4\times 10^{-5}\, M_\odot\,\pc^{-3}$, with an uncertainty of about
50\%.
The power-law indices $\alpha=0.25$ for the spheroid and
$\alpha=0.44$ for the disk (both uncorrected for binaries)
are similar to those of
globular clusters of moderate to high metallicity.

Subject Headings:  stars: low mass, luminosity function
\smallskip

\endpage

\chapter{Introduction}

	Subdwarfs comprise the great majority of stars in the Galaxy's
spheroidal component.  There are three main reasons to study the luminosity
function (LF) and physical distribution of these objects.  

	First, microlensing
results indicate that a substantial fraction of the Galaxy's dark matter
may be in compact objects (Alcock et al.\ 1997).  While spheroid stars
themselves certainly cannot be responsible for the majority of the microlensing
events (Bahcall et al.\ 1994, hereafter Paper I; Graff \& Freese 1996; 
Reid et al.\ 1996; Flynn, Gould, \& Bahcall 1996, hereafter Paper II; 
M\'endez et al.\ 1997), it
is possible that substellar objects in the spheroid do make a non-negligible
contribution.  The {\it shape} of the spheroid stellar
LF and hence the shape of its stellar
mass function (MF) provide an important clue by extrapolation
to the density of 
these substellar spheroid objects (M\'era, Chabrier, \& Schaeffer 1996).

	Second, by comparing the spheroid LF with that of globular clusters, 
one can gain insight into the evolution of the latter.  Globular clusters 
appear to have
anomalously low mass-to-light ratios compared to other old systems with
dynamically measured masses such as elliptical galaxies and the bulges of
spirals.  A plausible explanation for this discrepancy is that the globulars
have lost the majority of their initial mass by evaporation of their low-mass
stars.  If this explanation were correct, then one would expect the LF and MF
of field stars to be rising more steeply toward low masses than the LFs and
MFs of globular clusters.

	Third, spheroid stars are an unwanted foreground in studies of
extra-galactic objects, such as counts of faint galaxies.  An accurate estimate
of the spheroid stellar density is useful both for planning observations and
for removal of this background (Bahcall 1986).

	There are two basic approaches for determining the spheroid LF.
The first, pioneered by Schmidt (1975), is to extract a local sample of
spheroid stars from a proper-motion catalog, measure their parallaxes (and
so their absolute magnitudes), and then estimate their density
as a function of absolute magnitude.   To avoid contamination by disk stars
which are more numerous than spheroid stars in the solar neighborhood, it is
necessary to set stringent kinematic selection criteria (Bahcall \& Casertano
1986, hereafter BC).  
These criteria must then be properly modeled in order to extract
the underlying LF from the observed counts.  
BC applied this method to 94 stars with transverse speeds $V_T>220\,\kms$
taken from the Eggen (1979a, 1980) proper motion survey.  The distances
were determined photometrically based on Eggen's (1979b) linear color-magnitude
relation.  Dahn et al.\ (1995, hereafter
DLHG) applied essentially
the same method to a sample of 114 stars with $V_T>220\,\kms$
taken from the Luyton (1979) 
proper motion catalog and for which they obtained reliable trigonometric 
parallaxes.  Because trigonometric parallaxes are fundamentally more reliable
than photometric parallaxes and because the actual color-magnitude relation is
neither linear nor one-to-one (Baraffe et al.\ 1997; and \S\ 3.2, below) 
we compare our results primarily to DLHG.
The DLHG LF peaks near $M_V=11.5$, similar to the peak of 
the disk LF (Stobie, Ishida \& Peacock 1989; Kroupa, Tout \& Gilmore 1993; 
Reid, Hawley, 
\& Gizis 1995; Gould, Bahcall, \& Flynn 1996, 1997 -- hereafter Papers III and 
IV.)

	An alternative method is to determine the spheroid LF from star 
counts.  The major difficulty of this approach has been that stars
could be reliably distinguished from galaxies only to relatively bright
magnitude limits, typically $V\lsim 20$.  At these magnitudes and for most
colors, disk stars greatly outnumber spheroid stars and it is therefore 
difficult to isolate a spheroid sample.  For this reason, Bahcall \& 
Soneira (1980), when they first applied this method, restricted attention to 
blue
stars near the main-sequence turn-off which are relatively isolated from
the disk population in the color magnitude diagrams (CMDs)
of the deepest ground-based
images of the time.  Richer \& Fahlman (1992) extended this approach to 
redder subdwarfs by counting stars in a pair of deep CCD images at high
Galactic latitude.  They reported a LF that is steeply rising at faint 
magnitudes in sharp contrast to the LF of DLHG which falls in
the same region.  However, the faint end $(V-I>1.75)$
of the Richer \& Fahlman (1992) 
spheroid sample is severely contaminated with disk stars (Reid et al.\ 1996).

	Here we analyze star counts from 53 fields imaged with the Wide Field
Camera (WFC2) on the repaired {\it Hubble Space Telescope (HST)}, covering a 
total area of
$221\,\rm arcmin^{2}$.  One can unambiguously distinguish stars from
galaxies in these fields to a mean limiting magnitude $I=23.8$, several 
magnitudes fainter than is possible from the ground.  This faint limiting
magnitude provides two major 
advantages relative to ground-based measurements.  First, one can measure 
the vertical profile of disk M stars and thereby determine the minimum 
magnitude (as a function of color) beyond which disk stars cease to be a 
serious contaminant.  By establishing a ``disk-free'' threshold, one 
eliminates the largest potential source of systematic error, contamination
at the red end by disk stars.  One could in principle measure the vertical 
profiles of late G or K stars from the ground since these are substantially
brighter.  These profiles should be similar to that of M stars.  However,
for these earlier type stars, one risks contamination from evolved spheroid
stars (Paper III).
The second advantage is that one can search for spheroid stars for 
several magnitudes beyond this disk-free threshold, allowing one to determine
the three-dimensional distribution of spheroid main-sequence stars for the 
first time.

	We derive in this paper the spheroid LF over the range 
$6.5<M_V<14.5$.   The LF is relatively flat or slightly rising over this range,
in contrast to the spheroid LF of  DLHG which shows a distinct peak at
$M_V\sim 12$ and also in contrast to several recently measured globular 
cluster LFs which peak near $M_V\sim 10$.  We also derive a MF, which shows no
obvious structure. We fit the MF to a power law $d N/d \ln M\propto M^\alpha$,
and find $\alpha=0.25 \pm 0.32$.  We derive an empirical color-magnitude
relation in order to be able to extract a LF from the photometric data.
The MF should be interpreted more cautiously than the LF, since
to extract a MF from photometric data one requires 
mass-luminosity and mass-color relations.  While empirical mass-luminosity
relations are available for disk stars (Henry \& McCarthy 1993), none have
been established for the spheroid.  Hence, we rely on the purely 
theoretical calculations of Baraffe et al.\ (1997) for the mass-luminosity
relation.  

	In \S\ 2, we review the observations and data reduction.  In
\S\ 3, we discuss our parameterization of the spheroid and our construction
of a color-magnitude relation.   In \S\ 4, we extract the LF and MF from the
data, and in \S\ 5, we discuss some of the implications of these results.

\chapter{Observations and Data Reduction}

\FIG\one{
Dereddened CMD of 166 spheroid stars detected in 53 fields
observed with the {\it HST} WFC2.  The Groth Strip fields ({\it circles}) and
other fields ({\it crosses}) are shown separately.  The vertical line at the
left is the color selection criterion $(V-I)_0\geq 1.07$.  The diagonal line
is the criterion excluding disk stars within 8 kpc of the Galactic plane and
evaluated for $b=60^\circ$ (appropriate for the Groth Strip).  The horizontal
line is the approximate magnitude limit for 27 of the 28 Groth Strip fields.
}
\topinsert
\mongofigure{ps.fig1x}{6.4}{5.5}{6.4}{
Dereddened CMD of 166 spheroid stars detected in 53 fields
observed with the {\it HST} WFC2.  The Groth Strip fields ({\it circles}) and
other fields ({\it crosses}) are shown separately.  The vertical line at the
left is the color selection criterion $(V-I)_0\geq 1.07$.  The diagonal line
is the criterion excluding disk stars within 8 kpc of the Galactic plane and
evaluated for $b=60^\circ$ (appropriate for the Groth Strip).  The horizontal
line is the approximate magnitude limit for 27 of the 28 Groth Strip fields.
}
\endinsert

	The sample is selected from the stars found in 53 fields with a 
total area of $221\,\rm arcmin^2$ imaged with 
WFC2 on {\it HST}.  The field
centers and limiting magnitudes are given in Table 1 of Paper IV.  The
procedure for identifying stars and measuring their fluxes is summarized
in Paper IV which refers to Papers I, II, and III for further details.

	Figure \one\ shows the $I$, $V-I$ CMD for the
total of 166 stars that meet the two selection criteria described below.
	The selection criteria were devised so as to obtain a nearly pure
sample of spheroid subdwarfs.  First, we exclude disk dwarfs by demanding that 
the inferred distance from the Galactic plane (assuming a {\it disk} 
color-magnitude relation: $M_V = 3.37(V-I)+2.89$) be at least 8 kpc.
In Paper IV, we measured the vertical profile of disk stars (including both
the thin-disk and intermediate populations).  From Figure 1 of Paper IV, it
is clear that few disk stars have inferred distances above 6 kpc.  We 
nevertheless adopt a still more conservative limit of 8 kpc because disk stars 
are $\sim 10^3$ times
more common than spheroid stars at the plane and so could be a 
serious contaminant even at relatively low densities.  The disk stars need
not actually be above 8 kpc to cause contamination: the intermediate disk
population is more metal-weak and hence less luminous than the main disk 
population, so that the true distances may be as little as half the inferred
distances for the most distant stars.  Nevertheless, since it is the 
{\it inferred distances} that are shown in Figure 1 of Paper IV, the 
8 kpc {\it inferred-distance} cutoff will remove essentially all disk stars.
The diagonal line
in Figure \one\ shows this threshold for the Galactic latitude $b=60^\circ$,
the value for the 28 Groth Strip fields.  Note that the detected stars
are not bunched up against this threshold as they would be if the sample
were contaminated by disk stars.

Second, we exclude spheroid turn-off stars and giants by restricting attention
to stars with $(V-I)_0\geq 1.07$.  This color cutoff 
eliminates turn-off stars since these
are bluer than the cutoff.  Metal-poor giants do exist with $V-I>1.07$ and 
these would remain in the sample if the 
color inequality were the only selection 
criterion.  However, these metal-poor giants have absolute magnitudes
$M_V<0.6$, which is more than $5.9$ mag brighter than disk stars of the same
color.  Thus, any giant satisfying both criteria would have to lie more than
 $120\,$kpc from the Galactic plane, where the density of giants is extremely 
small.  Explicitly, the fraction of giant contaminants in a given 
apparent-magnitude interval is
$$ {N_{\rm G}\over N_{\rm MS}} = {\nu (D_{\rm G},l,b)\over\nu (D_{\rm MS},l,b)}
\,\biggl({D_{\rm G}\over D_{\rm MS}}\biggr)^3\,
{\Phi_{\rm G}\over \Phi_{\rm MS}}\sim {\Phi_{\rm G}\over \Phi_{\rm MS}}
\eqn\gaintmscomp$$
where $\nu (D_{\rm G},l,b)$ is the density of giants (relative to their local
density) at their distance
$D_{\rm G}$ and Galactic coordinates $(l,b)$, $\Phi_{\rm G}$ is the local
normalization of the giant LF at their absolute
magnitude (inferred from their color),
and the corresponding quantities for main-sequence stars are similarly defined.
The last step follows because both the giants and main-sequence stars are
sufficiently far that their Galactocentric distances $R$ are of the same
order as their distances from us, $D$.  Since $\nu\sim R^{-3}$, the two
terms approximately cancel.  
Since ${\Phi_{\rm G}/\Phi_{\rm MS}}\sim {\cal O}(1\%)$, 
it follows that giant contamination is negligible.

Another potential contaminant is disk white 
dwarfs.  Old white dwarfs could be seen to a distance of $1\,$kpc and younger
WDs could be seen even further.  However, using the local disk white dwarf LF 
of Liebert, Dahn, \& Monet (1988) and the vertical disk profile for M dwarfs
reported in Paper IV, we find that $<1$ WD is expected in the 53 fields 
combined.  White dwarfs should have a vertical profile like the M dwarfs
because their main-sequence progenitors have a mixture of ages that is similar
to that of M dwarfs.  

	For completeness, we also consider spheroid white dwarfs.  As we
show in \S\ 5.1, at the Galactic plane the spheroid has only $\sim 1/600$
of the density of disk.  At 1 kpc above the plane, the edge of the volume 
where the peak of the white dwarf LF is visible, this fraction is 
$\sim 1/60$.  Hence, spheroid white dwarf contamination is almost two orders
of magnitude smaller than that caused by disk white dwarfs, and thus completely
negligible.  Finally, disk giants are far too bright to enter the sample.  

	QSOs (or AGN of lower luminosity) are another possible source of
contamination.  The density of QSOs with $B<22$ is $\sim 200\,\rm deg^{-2}$
(Hartwick \& Schade 1990), implying that a total of $\sim 12$ QSOs should be 
present in our fields.  While the QSO LF is not known beyond $B=22$, one might
plausibly assume that the number continues to double with each magnitude.
Since the survey extends approximately 3 magnitudes beyond this limit, there
could be ${\cal O}(100)$ such objects in the 53 fields.  The great majority of
these QSOs are too blue to pass the color selection criterion of 
$(V-I)_0\geq 1.07$.
For example, we obtained ground-based $V$ and $I$ photometry of 115 QSOs in the
course of measuring the colors of stars found in pre-repair {\it HST} images 
(Paper III).
Only 7 of these 115 have $(V-I)_0\geq 1.07$.  (These have 1950 coordinates and
corresponding redshifts: 
0438$-$43, 2.852; 
0846$+$51, 1.860; 
0903$+$17, 2.771; 
1011$+$09, 2.260; 
2121$+$05, 1.878; 
2136$+$14, 2.427; and 
2225$-$05, 1.981). 
QSOs should exceed this color limit only
if they are at $z>4$ or have substantial internal extinction.  In addition,
most AGN are embedded in discernible galaxies.  The lower the AGN
luminosity, the more likely it is that the diffuse light of the host
galaxy will cause the object to be rejected as ``non-stellar'' in our 
initial morphological selection.  There are no data from which one could
measure the rejection fraction precisely, but the one available piece of
evidence is encouraging: a $V\sim 25$, $z=3.368$ emission-line galaxy was
identified by two groups in the Hubble Deep Field (HDF) and characterized
as ``point-source?'' by one 
(Steidel et al.\ 1996, object C2-11) and ``[d]espite some faint extended
emission [has the] smallest half-light radius in our sample, 
$r_{1/2}=0.\hskip-2pt ''14$,
indistinguishable from a point source'' by the other (Lowenthal et al.\ 
1997, object hd2\_0705\_1366).
However, in our analysis of HDF (1996, Paper II) we classified this object as
``non-stellar'' while noting that it is compact.  

	For the reasons given in the preceding paragraph, we believe that 
our subdwarf sample is not significantly contaminated by
compact extra-galactic objects, and we assume no contamination in the analysis
below.  This assumption could be tested by searching for QSOs
(using either objective-prism or broad-band techniques) in
 the Large Multi-Color Survey (``Groth Strip'') that comprises 28 of
the 53 fields analyzed here.  There should be $\sim 6$ QSOs in these fields
with $B<22$.  If these were rejected as ``non-stellar'' in our morphological
selection, it would indicate that contamination is indeed minor.  In addition,
such a study would provide valuable data on the host environments of faint
QSOs that would be complementary to the studies by Bahcall et al.\ (1997)
and Jones et al.\ (1997) on the hosts of bright QSOs.

	We conclude that
the sample of 166 spheroid stars is nearly free of contamination.  We believe
that no more than one, or perhaps a few, members of the sample
are objects other than spheroid subdwarfs.

\chapter{Characterization of the Spheroid}

\section{Spheroid Parameterization}

	We model the distribution of spheroid stars as functions of Galactic
coordinates $(x,y,z)$ and absolute magnitude $M_V$ by a flattened power law,
$$n(x,y,z;M_V,c,\ell,R_0) = \Phi(M_V)\nu(x,y,z;c,\ell,R_0),\eqn\flatspher$$
where $\Phi(M_V)$ is the local LF, $\nu$ is the density of the spheroid as
a function of position normalized to the solar neighborhood,
$$\nu(x,y,z;c,\ell,R_0) = 
\biggl[{x^2 + y^2 + (z/c)^2\over R_0^2}\biggr]^{-\ell/2},\eqn\nudef$$
$R_0$ is the galactocentric distance, $c$ is the flattening parameter, 
and $\ell$ is the power.  Thus there are three free galactic-structure
parameters $(c,\ell,R_0)$, plus one free parameter for each luminosity bin.  

\section{Color-Magnitude Relation}

	In order to interpret the observables $(I,V-I)$ in terms of the
parameters of the model $(\Phi(M_V),c,\ell,R_0)$, one must assume a 
color-magnitude relation (CMR).  For most globular clusters, the main sequence
forms a narrow line with very little scatter, and the CMR is a tight 
one-to-one relation between color
and absolute magnitude.  By contrast, the spheroid is composed of stars
with a wide range of metallicities and thus a range of absolute magnitudes
at fixed color.  Hence, the one-to-one CMR relation characteristic of
globulars must be replaced by a probability distribution.  To calibrate this
relation, we rely primarily on the CMD of nearby 
subdwarfs with transverse speeds $V_T>260\,\kms$ kindly provided to us in 
advance of publication by C.\ Dahn (1997 private communication).  This CMD
is updated from the work of DLHG and we therefore refer to it as the
``DLHG CMD'' or ``DLHG stars''.  The high velocity DLHG
stars should be representative of the stars in our sample, which
are found many kpc from the Galactic plane (see below).  
Unfortunately, the 43 DLHG subdwarfs with $V_T>260\,\kms$ and
reliable parallaxes do not sample the CMD densely enough to permit 
direct construction of a CMR.  We therefore 
use the low-metallicity theoretical isochrones of Baraffe et al.\ (1997) 
to interpolate across the DLHG CMD.  We proceed as follows:
First, we superpose the isochrones (at [m/H]$=-2.0,$ $-1.5,$ $-1.3,$ and
$-1.0$) on the DLHG CMD.  We find, as was already noted by
Baraffe et al.\ (1997), that many of the stars are brighter than even
the most metal-rich of these isochrones.  We therefore add an additional
isochrone at [m/H]$=-0.5$ kindly provided to us by I.\ Baraffe (1997 private
communication).  We then estimate the [m/H] of each of the
43 stars by interpolating between the isochrones.  We find that the cumulative
metallicity distribution, $N(\rm [m/H])$, is well represented by
$ {d N/ d\rm [m/H]} = 0.49$ for $-1.8<{\rm [m/H]}<-0.9$, and
$ {d N/ d\rm [m/H]} = 
0.93$ for $-0.9<{\rm [m/H]}<-0.3$,
with a mean metallicity $\VEV{[\rm m/H]}=-0.93$.
This is somewhat surprising because high-velocity spheroid stars are generally
believed to be more metal poor, $\VEV{\rm [Fe/H]}\sim -1.5$ 
(Laird et al.\ 1988; Nissen \& Schuster 1991).  Part of the
difference, perhaps 0.35 dex, can be accounted for by the fact that Population
II stars are more deficient in Fe than in metals generally.  The remainder
could in principle be a result either of a previous error in estimating the 
mean abundances of spheroid stars or of problems with the theoretical 
isochrones.  Gizis (1997)  measured the metallicities of a subset of the
high-velocity DLHG stars and found that they lay in the traditional
normal range for metal-poor stars, [Fe/H]$\sim -1.5$.  He noted, as we have,
that the Baraffe et al.\ (1997) isochrones appear to be inconsistent with
this metallicity and pointed out that the previous generation of the same code 
(Baraffe et al.\ 1995) seemed to yield closer agreement with his 
spectroscopically measured metallicities.  On the other hand,
I.\ Baraffe (1998, private communication) and her collaborators believe that
the new codes are superior in that they take account of more of the physics
and show better agreement with globular-cluster data.  Thus, the issue remains
unresolved.  Fortunately, for present purposes, 
the discrepancy is not a concern because we use the
theoretical isochrones only to interpolate between the data points.  However,
as we discuss below, this discrepancy will be of greater concern when we
estimate the subdwarf MF.

\section{Kinematic Versus Photometric Selection}

	The DLHG stars were selected according to kinematic
criteria and therefore could in principle be biased relative to the 
photometrically selected {\it HST} sample.  Suppose that the spheroid were
composed of sub-populations and that those populations with a larger 
asymmetric drift relative to the Local Standard of Rest, $v_a$,
also had lower metallicities.  The DLHG sample that we use to calibrate the
CMR is selected from
stars with transverse velocity $V_T>260\,\kms$, so the stars with
lower metallicity would be over-represented.  Since these are more subluminous,
we would tend to underestimate the luminosities and hence the distances of the
stars in our sample.  We now argue that this
bias is likely to be small on the basis of two complementary arguments.

	First, two independent estimates of local spheroid kinematics find very
similar results.  Casertano, Ratnatunga, \& Bahcall (1990, hereafter CRB) 
find dispersions
$(\sigma_R,\sigma_\phi,\sigma_z) = (160,89,94)\,\kms$ and
$v_a=217\,\kms$ from samples of
high proper motion stars.  Layden et al.\ (1996) find
$(\sigma_R,\sigma_\phi,\sigma_z) = (168,102,97)\,\kms$ and $v_a=198\,\kms$
using spheroid RR Lyrae stars whose selection depends only weakly on
kinematic criteria.  The spheroid is therefore rotating very slowly.
If the spheroid is composed of subpopulations, each subpopulation
is probably also rotating slowly.  Otherwise, some subpopulations would have
to be counter-rotating.  The difference in bulk velocity between
populations should then be no more than a few tens of $\kms$.  
We find numerically that the selection function for the DLHG stars
for $v_a=230\,\kms$ is only $\sim 30\%$ higher than
for $v_a=200\,\kms$.  Even if the entire dispersion in the DLHG CMD of
$\sigma\sim 0.4\,$mag is due to metallicity variation that is perfectly
correlated with asymmetric drift, this implies that the bias toward 
underestimating the luminosity of the stars in our sample is only
$\sim 0.1\,$mag.

	Second, Beers \& Sommer-Larson (1995) have used radial velocities to
measure the asymmetric
drift of a non-kinematically selected sample of metal-poor stars as
a function of metallicity.  They find that for [Fe/H]$<-1.5$, the asymmetric
drift is constant (see their Fig.\ 6).  For more metal-rich stars,
there is a strong dependence on metallicity.  Chiba \& Yoshii (1998)
find a similar result from a sample of metal-poor giants and 
RR Lyrae stars with 
Hipparcos proper motions (see their Fig.\ 5).
The simplest interpretation
of these results is that the stars with [Fe/H]$<-1.5$ are drawn almost
entirely from the spheroid, and that the spheroid has no differential
rotation.  The more metal-rich parts of this sample are increasingly 
contaminated with disk or thick disk stars.  In brief, we believe that we 
and DLHG are sampling essentially the same population.

\chapter{Analysis}

\section{Properties of the Sample}

\FIG\two{
Galactic coordinates $(\rho,z)$ of 166 spheroid stars ({\it crosses})
found in 53 {\it HST} WFC2 fields, together with the most distant coordinates
probed by each field ({\it circles}).  Note that the circles {\it do not}
represent detected stars.  Here $\rho^2 = x^2+y^2$ and $\rho$
is defined to have the same sign as $x$.  The Galactic center is shown as a
``+'' and the Sun's position is shown as a ``$\odot$''.  The two horizontal
lines (at $z=\pm 2.34$ kpc) represent the exclusion of disk stars within 
8 kpc of the Galactic Plane.  The apparent discrepancy between these two
distances arises because spheroid stars can be as much as 
$5\log (8/2.34)=2.67\,$mag fainter than disk stars at the same color.
All distances are evaluated using the 
[m/H]$=-1.0$ isochrone of
Baraffe et al.\ (1997).  The dense ``plume'' is the Groth Strip.
}
\topinsert
\mongofigure{ps.fig2x}{6.4}{5.5}{6.4}{
Galactic coordinates $(\rho,z)$ of 166 spheroid stars ({\it crosses})
found in 53 {\it HST} WFC2 fields, together with the most distant coordinates
probed by each field ({\it circles}).  Note that the circles {\it do not}
represent detected stars.  Here $\rho^2 = x^2+y^2$ and $\rho$
is defined to have the same sign as $x$.  The Galactic center is shown as a
``+'' and the Sun's position is shown as a ``$\odot$''.  The two horizontal
lines (at $z=\pm 2.34$ kpc) represent the exclusion of disk stars within 
8 kpc of the Galactic Plane.  The apparent discrepancy between these two
distances arises because spheroid stars can be as much as 
$5\log (8/2.34)=2.67\,$mag fainter than disk stars at the same color.
All distances are evaluated using the 
[m/H]$=-1.0$ isochrone of
Baraffe et al.\ (1997).  The dense ``plume'' is the Groth Strip.
}
\endinsert

Figure \two\ shows the approximate positions ({\it crosses}) 
of the 166 stars in cylindrical coordinates
$(\rho,z)$ where $\rho^2\equiv x^2+y^2$.  The distances are determined from
the measured colors and apparent magnitudes, and assuming the 
color-magnitude relation for the [m/H]$=-1.0$ isochrone of Baraffe et al.\ 
(1997).  Also shown is the minimum distance from the plane (2.34 kpc)
that spheroid stars could have been detected 
due to the exclusion of disk stars within 8 kpc of the
Galactic plane ({\it solid lines}).  The apparent discrepancy between these
two distances is due to the fact that spheroid stars can be as much as
$5\log (8/2.34)=2.67\,$mag fainter than disk stars at the same color.
The maximum distances probed for each of the 53 fields are shown as
({\it circles}).  Note that these circles {\it do not} represent detected
stars.  The densely populated ``plume'' is the Groth Strip.

	Note that the spheroid is well sampled in several substantially
different directions out to Galactocentric
distances of $R\sim 20\,$kpc and that most lines of sight probe to
$R\sim 40\,$kpc (even though there are relatively few detections at these
large distances).  These characteristics give good leverage on the
Galactic structure flattening parameter, $c$ and the power law, $\ell$.
The fact that some lines of sight extend to negative $x$ values (shown as
negative $\rho$ in Fig.\ \two) implies that the sample should give modest
leverage on $R_0$.  Note that the most distant star detected has a 
galactocentric distance $R\sim 45\,\kpc$.

\section{Likelihood Function}

	Let $\tau_{i j k}$ be the expected number of stars in the 
bin of apparent magnitude $I_i$, color $(V-I)_j$, for the
$k$th field.  The Poisson probability of finding $n_{i j k}$ stars in this 
bin is then $P_{i j k} = \exp(-\tau)\tau^n/n!$.  If the bins are chosen
to be very small so that $\tau\ll 1$, then  $n=0$ or $n=1$, so
$n!\rightarrow 1$.  Hence, the logarithm of the likelihood is
$$ \ln L = \sum_{i,j,k} \ln P_{i j k} = 
\sum_{i,j,k}n_{i j k}\ln(\tau_{i j k}) - \sum_{i,j,k} \tau_{i j k}.\eqn\lnlik$$
The second term on the right hand side is simply $N_{\rm exp}$, the expected
total number of stars to be detected for the model, 
while the first reduces to a sum over
the detected stars:
$$ \ln L = \sum_{{\rm det},i,j,k}\ln\tau_{i j k} - N_{\rm exp}
.\eqn\lnliktwo$$
To maximize $\ln L$ over the class of models represented by equation
\flatspher, we need to predict $\tau_{i j k}$ as a function of Galactic
parameters.  We first evaluate $\CMD_{i' j' l}$, 
the color-magnitude distribution
 in $M_I$ and $(V-I)_0$ (binned by indices $i'$ and $j'$) of stars uniformly
distributed over the $l$th bin of absolute magnitude $M_V$, and 
distributed in metallicity as described in \S\ 3.2.  For each 
distance-modulus bin $\mu_m$, we then construct a normalized 
color-apparent magnitude diagram $\cmd_{i j k l m}$ by first translating
$\CMD_{i' j' l}$ by $\mu_m + A_{I,k}$ in the magnitude direction and $E_k(V-I)$
in the color direction, and then convolving with the observational errors.
We define the local volume element 
$$V_{k,m} = {\ln 10\over 5} 10^{0.6\mu_m + 3}\Omega_k \Delta\mu\ \pc^3,
\eqn\volele$$
where $\Omega_k$ is the angular area of the $k$th field (see Paper IV) and
$\Delta \mu$ is the width of the distance modulus bins.  This allows us
to write the first term in equation \lnliktwo\ as
$$ \sum_{{\rm det},i,j,k} \ln\tau_{i j k}
= \sum_{n=1}^{N_{\rm det}} \ln \sum_{l,m}
\Phi_l \nu_{k(n),m}(c,\ell,R_0) V_{k(n),m}\cmd_{i(n),j(n),k(n),l m},
\eqn\firstterm$$
where $N_{\rm det}$ is the total number of stars detected in all fields, and
$\nu$ is the Galactic structure function given by equation \flatspher.  
We  adopt bin sizes of 0.1 mag for the magnitude indices
over which we integrate
($i,i',$ and $m$) and 0.025 magnitudes for the color indices ($j$ and $j'$).
Similarly, we write the second term in equation \lnliktwo\ as
$$ N_{\rm exp} = \sum_{k,l,m}\Phi_l \nu_{k m}(c,\ell,R_0)V_{k m}
\cmd_{{\rm tot},k l m}, \eqn\secondterm$$
where
$$\cmd_{{\rm tot},k l m}\equiv
\sum_{{\rm selection},i,j}\cmd_{i j k l m},\eqn\cmdtot$$
and where the sum is restricted to the portions of the CMD
satisfying the selection criteria.  The matrices 
$\cmd_{i(n),j(n),k(n),l m}$ and $\cmd_{{\rm tot},k l m}$ can be evaluated
in about 15 minutes on a SPARC 5.  Once these are determined, the likelihood
function and its derivatives with respect to all the parameters can be
evaluated in about 1 second, and hence parameter space can be 
explored rapidly.

\section{Luminosity Function}

\subsection{4.3.1 Best-Fit LF}

	We use the formalism of the previous section to evaluate simultaneously
the Galactic structure parameters $c,\ell,$ and $R_0$ and the LF with the
latter being broken up into four 
2-mag bins centered at $M_V=7.5,$ 9.5, 11.5, and 13.5.  
We note that the full range of the LF must be chosen broad enough so that
no stars in the detected color range $1.07\leq (V-I)_0\leq 2.52$ could have
absolute magnitudes outside the range of the LF.  
Otherwise, any such stars that are
detected will be falsely attributed by the likelihood function to stars within
the range, and the LF will be overestimated.  The adopted limits satisfy this
criterion.  On the other hand, there is
no systematic tendency to underestimate the LF if the end bins extend somewhat
beyond the color-selection range, since the likelihood function automatically
takes this selection into account.  
We find 
$$ c=0.96\pm 0.22,\qquad 
\ell = 2.96\pm 0.27,\qquad
R_0 = 6.2\pm 1.8\,\kpc,\eqn\parmsone$$ 
and LF (in units of $10^{-5}\,\pc^{-3}$) 
$\Phi(7.5) = 1.05\pm 0.55$,  
$\Phi(9.5) = 1.37\pm 0.64$,  
$\Phi(11.5) = 1.98\pm 0.78$,  and
$\Phi(13.5) = 1.64\pm 1.20$.  
While it is encouraging that the solution for $R_0$ in equation \parmsone\
is consistent with other determinations, our error bars are not
competitive with other methods of measuring the galactocentric distance.  
We henceforth fix $R_0=8\,\kpc$ (Reid 1993) in all further calculations.
We then find
$$ c=0.82\pm 0.13,\qquad 
\ell = 3.13\pm 0.23,\qquad
(R_0 \equiv 8.0\,\kpc)\eqn\parmstwo$$
and a LF which is similar in shape to the one obtained without fixing $R_0$, 
but is about 15\% smaller in normalization:
$\Phi(7.5) = 0.85\pm 0.36$,  
$\Phi(9.5) = 1.12\pm 0.40$,  
$\Phi(11.5) = 1.72\pm 0.57$,  and
$\Phi(13.5) = 1.46\pm 1.05$.  

\FIG\three{
Luminosity functions of the spheroid as determined in this paper using {\it
HST} data
({\it filled circles}) compared to that of DLHG ({\it squares}) and
BC ({\it open circles}).  Also shown is the average 
LF of three metal-poor globular clusters ({\it stars}) as measured by 
Piotto et al.\ (1997).  The DHLG and BC LFs have been multiplied by 0.75 and
0.62 respectively based on a reanalysis of spheroid kinematics by CRB.  
The globular cluster LF is 
normalized arbitrarily.  Two sets of error bars are shown for the {\it HST}
LF.  The smaller errors assume that the Galactic structure parameters are
fixed.  The pattern of these error bars shows that the LF of brighter stars
is much more sensitive to assumptions about Galactic structure compared to
the LF of fainter stars.  To avoid clutter, no error bars are shown for the
BC LF.  In addition, the last point of the BC LF (at $M_V=12.5$) is not 
shown because it is based on only two stars.
}
\topinsert
\mongofigure{ps.fig3x}{6.4}{5.5}{6.4}{
Luminosity functions of the spheroid as determined in this paper using {\it
HST} data
({\it filled circles}) compared to that of DLHG ({\it squares}) and
BC ({\it open circles}).  Also shown is the average 
LF of three metal-poor globular clusters ({\it stars}) as measured by 
Piotto et al.\ (1997).  The DHLG and BC LFs have been multiplied by 0.75 and
0.62 respectively based on a reanalysis of spheroid kinematics by CRB.  
The globular cluster LF is 
normalized arbitrarily.  Two sets of error bars are shown for the {\it HST}
LF.  The smaller errors assume that the Galactic structure parameters are
fixed.  The pattern of these error bars shows that the LF of brighter stars
is much more sensitive to assumptions about Galactic structure compared to
the LF of fainter stars.  To avoid clutter, no error bars are shown for the
BC LF.  In addition, the last point of the BC LF (at $M_V=12.5$) is not 
shown because it is based on only two stars.
}
\endinsert

   	Figure \three\ shows the LF derived in this paper
along with the local spheroid LFs of DLHG and BC and an
average LF of three metal-poor globular clusters (NGC 6341, NGC 7078, and
NGC 7099) measured by Piotto, Cool, \& King (1997).  
The DLHG and BC LFs are multiplied by a factor 0.75 and 0.62 respectively
(as discussed below), and the cluster LF is arbitrarily normalized.  

The {\it HST}
spheroid LF is shown with two sets of error bars. One set of errors is 
obtained as described above.
The other (smaller) error bars are determined by fixing the Galactic structure
parameters $c$ and $\ell$ at their best-fitting values.  Note that the 
difference is dramatic for the brightest bin but noticeable only with a
magnifying glass for the 
faintest bin.  This is because the brighter stars probe distant regions of
the Galaxy and hence their LF is highly correlated with the Galactic structure
parameters.  By contrast, the fainter stars are relatively nearby and hence
insensitive to assumptions about the large-scale structure of the Galaxy.
Another feature of the {\it HST} LF, which is not illustrated in Figure \three,
is that the individual luminosity bins are anti-correlated with one another.
When $c$ and $\ell$ are held fixed, neighboring bins have correlation
coefficients of about $-0.3$.  This is due to the fact that most detected stars
can be almost equally well attributed to either of two neighboring luminosity
bins.  These various correlations among the parameters make
the interpretation of Figure \three\ less straight forward than one would like.

	The major question posed by Figure \three\ is: are the DLHG and 
{\it HST} 
LFs consistent?  Before addressing this question, we first justify our
reduction of the DLHG and BC LFs by a factor of 0.75 and 0.62 respectively.

\subsection{4.3.2 Spheroid Kinematics and the DLHG and BC LFs}

	BC selected stars with transverse speeds
$V_T>220\,\kms$ and assumed that the
underlying spheroid population had characteristics given by their two component
Galactic model, namely $v_a=-154\,\kms$ and 
$(\sigma_R,\sigma_\phi,\sigma_z) = (140,100,76)\,\kms$.  Based on this model,
they calculated that a fraction 0.33 of spheroid stars satisfied their
transverse-speed selection criterion.
Subsequently, CRB showed that a significantly better fit to the same data
can be obtained by assuming that there is a third population with
intermediate kinematics.  The spheroid component is then moving much more
rapidly relative to the Sun: $v_a=-217\,\kms$,
$(\sigma_R,\sigma_\phi,\sigma_z) = (160,89,94)\,\kms$ (CRB). As we discussed in
\S\ 3.1, this determination is in excellent agreement with the kinematics
of spheroid RR Lyrae stars as measured by Layden et al.\ (1996).  We have
therefore redone the calculation using BRC kinematics
and find a completeness factor 0.54.  Hence
we multiply the BC results by a factor 0.33/0.54 = 0.62 and label the modified
BC LF as ``BC/CRB''.  DLHG also selected stars with transverse speeds
$V_T>220\,\kms$.  They used slightly different kinematic assumptions and
derived a completeness factor of 1/2.46.  We therefore multiply their results
by a factor $1/(2.46\times 0.54)=0.75$ and label the resulting LF
``DLHG/CRB''.  

\subsection{4.3.3 Comparison of the {\it HST} and DLHG/CRB LFs}

	From Figure \three, one sees that the DLHG/CRB LF is overall higher
than the {\it HST} LF, 
particularly near the peak of the former, $M_V\sim 11.5$.
However, as we emphasized above, the
correlations among the parameters render difficult the 
interpretation of the figure.  The appropriate method to determine whether
these two measurements are consistent is to fix the {\it HST} LF at the 
DLHG/CRB values for the three overlapping bins ($M_V=9.5,$ 11.5, and 13.5) and
to allow the other parameters to vary.  We find that the best-fit such solution
has an increase in $\chi^2$ (i.e. $-2\ln L$), of 11.4 for three more degrees
of freedom.  This means that the two LFs differ at the $2.9\,\sigma$ level.
We note for completeness that this solution yields $c=0.638\pm 0.050$ and 
$\ell=3.27\pm 0.22$.  Of course, one might also solve for the LF that 
minimizes $\chi^2$ for the two samples simultaneously, rather than imposing
the DHLG/CRB solution of the {\it HST} data.  However, we find that the 
best-fit such solution is still discrepant by $2.8\,\sigma$.

	The LFs of the spheroid and the globular clusters cannot be 
directly compared because they are of different metallicities.  We reserve 
comparison for our discussion of MFs.

\subsection{4.3.4 Possible Explanations for the Discrepancy Between LFs}

	One possible reason for the discrepancy between between the {\it HST}
and DLHG/CRB LFs is a statistical fluctuation.  Assuming Gaussian statistics,
the chance of a $2.8\,\sigma$ event is $\sim 0.5\%$, but the probability
would rise rapidly if even a modest part of the difference between the LFs 
were due to unrecognized systematic errors in either determination.  One
indication of the possible
size of such systematic errors is the conflict between
the DLHG/CRB LF and the BC/CRB LF, both of which were based on local
proper-motion selected samples.  As we discussed in \S\ 1, the most likely
cause of this conflict is that the BC LF is based on Eggen's (1979b) linear, 
single-valued CMR while the DLHG LF is based on trigonometric parallaxes.
Hence, the DLHG solution is to be preferred a priori over that of BC.
Nevertheless, it is striking that the BC/CRB LF is actually in very good 
agreement with the {\it HST} LF.  In brief, we believe that no strong
conclusions can be drawn from the apparent conflict between the {\it HST}
and DLHG/CRB LFs.

	Just the same, it is worth asking if this difference could be a real
effect.  Sommer-Larsen \& Zhen (1990) have proposed that the spheroid has two
components, a highly flattened component which contributes about 40\% of the
total density in the neighborhood of the Sun, and a nearly spherical
component ($c=0.85\pm 0.12$) which contributes the other 60\%.  We emphasize
that the model's flattened component is supported by an anisotropic 
velocity dispersion tensor, in contrast to the more traditional (and also
highly flattened) intermediate or ``thick disk'' population which is supported
by rotation.  Sommer-Larsen \& Zhen (1990) developed this model based on 118
non-kinematically selected stars with [Fe/H]$\leq -1.5$.  Hartwick (1987)
advanced a similar notion on the basis of a study of RR Lyrae stars with
[Fe/H]$\leq -1.0$, but this sample is almost certainly contaminated with
intermediate population stars.

	If the Sommer-Larsen \& Zhen (1990) model were correct, then
only the spheroidal component would enter the {\it HST} sample.  The flattened
component would be effectively eliminated because the selection criteria 
remove all stars within several kpc of the plane (see Fig.\ \two).  It would 
then be appropriate to multiply the DLHG/CRB LF by 0.6 before comparing it to
the {\it HST} LF.  We find that $\chi^2$ then rises by only 3.6 for three
more degrees of freedom.  That is, the two LFs are consistent at the 
$1\,\sigma$ level.  Thus, the discrepancy between the {\it HST} and 
DLHG/CRB LFs could plausibly be explained by a two-component spheroid.
However, we caution that the evidence for a two-component spheroid is
limited, and it is quite possible that the discrepancy is due to a
combination of systematic and statistical errors.

	We note that a prediction of the two-component model is that
the velocities of spheroid stars perpendicular to the plane should contain
a hot and cold component, and therefore the velocity distribution should have 
a kurtosis in excess of the Gaussian value $K=3$.  Popowski \& Gould (1998)
find $K=4.5$ for 165 ``halo-3'' RR Lyrae stars.  For present purposes, it is
more appropriate to use the subsample of 97 stars restricted to [Fe/H]$<-1.5$.
For these we find $K=4.7$.  This is inconsistent with the Gaussian value
at the $2.7\,\sigma$ level.

\section{Mass Function}

	It is customary to determine the MF of a stellar population by
first measuring its LF and then converting to a MF using a mass-luminosity
relationship.  
However, the spheroid 
is composed of stars with a wide range of metallicities and hence a 
correspondingly wide range of masses at fixed luminosity; thus the usual
procedure for determining a MF is not applicable.  Moreover, the observables 
for our spheroid sample are color and flux (not luminosity) and there is no 
one-to-one relation between color and flux and either luminosity or mass.

	We adopt a different approach which is similar to the LF measurement 
that we described in \S\ 4.  We repeat for the mass function all the 
steps described in \S\ 4 except that we initially
construct $\CMD_{i' j' l}$ with $l$ running over {\it mass bins}, $M_l$, rather
than absolute magnitude bins as before.  That is, we draw stars uniformly
in log mass rather than log luminosity.  We use exactly the same models
from Baraffe et al.\ (1997) to do this with exactly the same metallicity 
distribution.  We stress that while this substitution is 
mathematically and computationally easy to perform, it contains strong
additional assumptions relative to the LF case.  For the LF, the Baraffe
et al.\ (1997) isochrones served only as interpolators between
the DLHG data points.  As such, systematic errors in the isochrones would
most probably not propagate into the analysis.  By contrast, the ``mass''
in these models is a purely theoretical quantity with no empirical calibration.
That is, the situation is very different than for the disk MF (Paper IV)
where an excellent empirical mass-luminosity relation exists (Henry \& McCarthy
1993).  Thus, the determination of the spheroid MF is on fundamentally weaker 
ground compared to the spheroid LF.

	We use the procedure just described
to evaluate the MF over the range $0.09< M/M_\odot<0.71$, the limits being
established according to the criterion outlined at the beginning of \S\ 4.3.1.
For four mass bins centered
at $(M/M_\odot)= 0.55,$ 0.33, 0.20, and 0.12, we find
$d N/d \log M = 14\pm 6$, $6\pm 4$, $12\pm 10$, and $19\pm 14$ in units
of $10^{-5}\,\pc^{-3}$.  The Galactic structure parameters are
$c= 0.80\pm 0.12$ and $\ell=3.15\pm 0.23$, i.e., almost identical to the 
values in the LF solution (eq.\ \parmstwo).   
This MF shows some hint of structure with a dip in the second bin, but one may
suspect that (as in the LF case) there is not actually enough information
in the data to resolve this structure.  

To test the information content of
the data, we fit them directly to a power-law mass function of the form
$${d N\over d\log M} = G(M;A,\alpha) =
A \biggl({M\over M_\odot}\biggr)^\alpha.\eqn\powerlaw$$
We modify the likelihood analysis discussed above in two ways.  First, we
calculate $\cmd_{i j k l m}$ for a large number of mass bins $M_l$
(in practice, $l= 1,..,16$).  Second, we write $N_{\rm exp}$ as
$$ N_{\rm exp}(c,\ell,R_0,A,\alpha) = 
\sum_{k,l,m}G(M_l;A,\alpha) \nu_{k m}(c,\ell,R_0)V_{k m}
\cmd_{{\rm tot},k l m}\Delta\log M,\eqn\nexpmass$$
where $\Delta \log M$ is the width of the logarithmic mass bin.  We also
write an analogous expression for the first term in equation \lnliktwo.
We find 
$$ c=0.79\pm 0.12,\qquad \ell =3.06\pm 0.22,\qquad (MF),\eqn\cellmassfun$$
and MF parameters $A=13.5\pm 7.4 \times 10^{-5}\,\pc^{-3}$ and 
$\alpha=0.25\pm 0.32$.  The error in the MF normalization, $A$, appears to be
extremely large but this is because the mass normalization ($M_\odot$) lies
outside the range of the data.  Hence, 
$A$ and $\alpha$ are highly correlated.
The correlation can be eliminated by normalizing to $0.225\,M_\odot$:
$${d N\over d\log M} = (9.4\pm 2.6)\times 10^{-5}
\biggl({M\over 0.225\, M_\odot}\biggr)^{0.25\pm 0.32}\,\pc^{-3}.\eqn\dndlnm$$

	The $\chi^2$ is 4.5 higher for the power-law solution compared to the
previous 4-bin solution, with 2 more degrees of freedom.  The binned
solution is therefore favored at the $1.6\,\sigma$ level, which could be
due to a statistical fluctuation,
systematic errors, or real structure in the MF.
In the absence
any compelling evidence for structure, we adopt the simpler power-law
parameterization given by equation \dndlnm\ as our best estimate of the MF.

\chapter{Discussion}

\section{Comparison of Galactic Structure Parameters}

	The best-fit Galactic-structure parameters in the LF and MF fits are
similar, ($c=0.79\pm 0.12,\ \ell =3.06\pm 0.22$)
and      ($c=0.82\pm 0.13,\ \ell = 3.13\pm 0.23$), respectively.  These
may be compared to previous determinations by various methods.

	Kinman, Wirtanen, \& Janes (1965) were the first to measure the
spheroid flattening.  They obtained $c=0.57$ from RR Lyrae stars.  Bahcall
(1986) found $c=0.80^{+0.20}_{-0.05}$ based on star counts.  Gilmore (1989) 
found $c$ to vary with Galactocentric radius, with $c\sim 0.5$ 
in the solar neighborhood, also based on star counts.
Most recently, Preston, Schectman, \& Beers (1991) also found $c$ to vary
with Galactocentric radius, but with $c\sim 0.7$ in the solar neighborhood
from RR Lyrae stars.

	Preston et al.\ (1991) measured the power-law $\ell=3.2\pm 0.1$
for RR Lyrae stars, and $\ell\sim 3.5$ for blue horizontal branch stars, the
latter being less well determined.  They
noted that these values were in good agreement with the value $\ell=3.5$
measured by both Harris (1976) and Zinn (1985) for globular clusters.

	Thus, the best-fit power-law found here is consistent with that
of RR Lyrae and blue horizontal branch stars.  The flattening is consistent
with the recent determination from RR Lyrae stars and with Bahcall's (1986)
measurement from star counts.  It is in conflict with Gilmore's (1989) 
determination at about the $2.5\,\sigma$ level.

\section{Comparison of Mass Functions}

\FIG\four{
MF for the spheroid ({\it bold line}) and the disk ({\it solid line})
as derived in this paper and Paper IV based on {\it HST} WFC2 data.  The
disk MF has been divided by 500.
}
\topinsert
\mongofigure{ps.fig4x}{6.4}{5.5}{6.4}{
MF for the spheroid ({\it bold line}) and the disk ({\it solid line})
as derived in this paper and Paper IV based on {\it HST} WFC2 data.  The
disk MF has been divided by 500.
}
\endinsert

	Figure \four\ compares the disk MF derived in Paper IV
with the spheroid MF derived here.  Neither is corrected for binaries.
In Paper IV, we argued that binaries should decrease the exponent of
(i.e., ``steepen'') the
disk MF by $\sim 0.35$ for $M<0.6\,M_\odot$, and so make the
right-hand part of the solid curve in Figure \four\  almost flat. 
We also argued that the correction for binaries 
should not affect the slope at the high-mass end.
To our knowledge, there are no data on the fraction of spheroid M
stars in binary systems 
and so we prefer to report the uncorrected result.  However, it may
be plausible to assume a similar correction for the disk and spheroid, in
which case the spheroid MF would also be approximately flat.  The uncorrected
spheroid and disk MFs have similar slopes and differ in normalization by
a factor $570\pm 160$ at the centroid of the spheroid determination, 
$M=0.225\,M_\odot$.

\FIG\five{
Power-law index ($\alpha$) versus metallicity ([m/H]) for the disk and
spheroid ({\it circles}) as determined from {\it HST} data and for
five groups of globular clusters ({\it triangles}) as determined by
Chabrier \&  M\'era (1997).  
}
\topinsert
\mongofigure{ps.fig5x}{6.4}{5.5}{6.4}{
Power-law index ($\alpha$) versus metallicity ([m/H]) for the disk and
spheroid ({\it circles}) as determined from {\it HST} data and for
five groups of globular clusters ({\it triangles}) as determined by
Chabrier \&  M\'era (1997).  
}
\endinsert

	In addition to comparing the spheroid MF to that of the 
disk, it is of interest to compare it to the MF of other metal-poor systems,
specifically globular clusters.  Capaccioli, Piotto, \& Stiavelli (1993)
have analyzed the MF slopes of 17 globular clusters over the range 
$0.5\,M_\odot\leq M \leq 0.8\,M_\odot$.  The slopes span 
a range $-1.2\lsim \alpha \lsim 1$ and show clear trends with galactocentric
radius and distance from the plane.  Capaccioli et al.\ (1993) argue that
this pattern confirms the prediction of Stiavelli et al.\ (1991) that all
clusters begin their life with the same MF (i.e., $\alpha\sim -1.2$) and
the clusters subject to the greatest dynamical effects preferentially lose
their low mass stars.  Since the spheroid MF is not affected by dynamics,
one would expect its MF to also have this slope, assuming the spheroid
and globular cluster MFs are similar.  This expectation is in strong conflict
with our result, $\alpha = 0.25\pm 0.32$.  However, it is possible that the
slope of the spheroid MF changes significantly at $M\sim 0.6\,M_\odot$
just as we have argued it does for the disk (Paper IV), in which case
the two MFs might still be consistent.  Capaccioli et al.\ (1993) also
analyzed some cluster MFs over the mass range $M\leq 0.4\,M_\odot$ but
regarded the mass-luminosity relations upon which they based their analysis
as unreliable.

	Chabrier \& M\'era (1997) have measured the MF of several globular
clusters by applying the Baraffe et al.\ (1997) isochrones (used 
to derive our spheroid MF in \S\ 4) to LFs from five groups of globular
clusters.  While the mass range varies from cluster to cluster, it extends
close to the bottom of the main sequence for most.  The most metal-poor 
of these groups ([m/H]$\sim -2.0$) is NGC 6341, NGC 7078, 
and NGC 7099, the same group whose LF is displayed in Figure \three.  The
most metal rich is 47 Tuc at [m/H]$\sim -0.5$.  The 
metallicities [m/H] and slopes $\alpha$ of the five groups are displayed in 
Figure \five\ as triangles.  The slopes of the spheroid MF (\S\ 4.4) and of
the disk MF (Paper IV) are shown as solid circles.  None of the determinations
are corrected for binaries.  The slopes of the disk and spheroid MFs are
consistent with the range set by the globular clusters of intermediate
to higher metallicities.  Only the three extreme low-metallicity clusters
at [m/H]$\sim -2.0$ have a slope substantially below (steeper than) this
range.  There is still controversy about the evaluation of globular cluster
MFs.  In particular, Piotto et al.\ (1997) find $\alpha\sim -1$ for the
three most metal-poor clusters compared to the value $\alpha=-0.5$ found
by Chabrier \& M\'era (1997).

\section{Mass Density of the Spheroid}

	In order to make a realistic estimate of
the mass density of the spheroid,
one must account for not only the detected objects 
($0.09< M/M_\odot < 0.71$)
but also those that for
one reason or another escape detection.  The latter include substellar objects
($M<0.09\,M_\odot$), upper main-sequence stars and evolved stars
($0.71<M/M_\odot\lsim 0.9$), remnants (which have progenitor masses
$M\gsim 0.9\,M_\odot$), and binary companions of the detected stars.  
In order to isolate the uncertainties
due to the last, we perform the entire calculation twice, first accepting
equation \dndlnm\ at face value and then correcting it for missing binaries.

	There are essentially no empirical data on substellar objects and very
little on higher-mass stars and remnants in the spheroid.  
We therefore make our 
estimates based on plausible, if highly debatable, assumptions.
For substellar objects, we assume that the power-law observed in the 
stellar-mass range continues into the brown dwarf regime to zero mass.
For the higher-mass stars and progenitors of remnants we assume a break
in the power-law to $\alpha=-1.7$ at the upper boundary of the observations,
that is,
$${d N\over d\log M} = 12.5\times 10^{-5}
\biggl({M\over 0.71\,M_\odot}\biggr)^{-1.7}\,\pc^{-3}
\qquad (M>0.71\,M_\odot).\eqn\dndlnmhigh$$

	We now justify this somewhat arbitrary estimate.
	There are only limited data constraining the slope of the MF
of metal-poor populations in the regime $M> 0.71\,M_\odot$.  BC have measured
the LF of spheroid turn-off stars ($0.7\lsim M/M_\odot\lsim 0.9$).  They find
(in units of $10^{-5}\,\pc^{-3}$ and after the correction discussed in \S\ 
3.2) of 
$\Phi(4.5) = 0.12\pm 0.12,$  
$\Phi(5.5) = 0.29\pm 0.14,$ 
$\Phi(6.5) = 0.93\pm 0.24,$ and
$\Phi(7.5) = 0.42\pm 0.10$.
For comparison,  we found $\Phi(7.5) = 0.85\pm 0.36$ in \S\ 4.3 (see Fig.\ 
\three).  In principle,
it would be possible to convert this LF to a MF and measure the slope.
In practice, the shortness of the baseline ($\Delta \log M \sim 0.1$) and the
size of the statistical errors make this impossible.  
An alternative approach would be to extend the log-mass baseline
by measuring the LF of spheroid white dwarfs.  By combining this LF with
white-dwarf cooling theory, one could hope to reconstruct the MF of the
white dwarf progenitors.  In fact, the white dwarf sample of Liebert et al.\
(1988) contains only 4 stars with transverse velocities $V_T>200\,\kms$.
These have $M_V=13.4,$ 13.6, 14.3, and 15.4, and so have progenitors of
mass $M\sim M_\odot$.  Hence, the baseline is again too short and the 
statistical fluctuations too large to determine the slope.  

	Another approach would be to adopt the MF measured for the upper
main sequence of globular clusters.  Recall from \S\ 5.2 that 
Capaccioli et al.\ (1993) measured a wide range of slopes for a collection
of 17 clusters in the mass interval $0.5\,M_\odot\leq M\leq 0.8\,M_\odot$, 
but argued that this variation was an artifact of dynamical effects.  They
concluded that the initial MF (the quantity most relevant to the spheroid MF
which does not suffer dynamical effects) is universal with $\alpha\sim -1.2$.
In fact, our adopted equation \dndlnmhigh\ has a similar slope.  However,
the main problem with all of these empirical estimates of Population II MFs
is that they apply only to a narrow range of masses below $1\,M_\odot$, while
the main contribution to the total mass comes from remnants of higher-mass
stars.

	We therefore investigate what is known about more metal-rich
populations and somewhat arbitrarily apply the results to the metal-poor
spheroid.  After a lengthy review of the available evidence (which seems
to indicate either a substantial variation in intermediate-mass MFs or 
substantial errors in their measurement)  Scalo (1998) says ``[i]f forced
to choose an IMF for use in galactic evolution studies,  I would suggest''
$\alpha = {-1.7\pm 0.5}$ for the range 
$(M_\odot < M < 10\,M_\odot)$.  The $\pm$ is intended to represent a
dispersion of measured values rather than an error.  Scalo (1998) recommends
a slightly shallower slope ($\alpha =-1.3$) for higher masses, but this 
change is
uncertain and has almost no impact on our estimate of the mass density.  For
simplicity, we therefore adopt $\alpha={-1.7}$ for
$M>M_\odot$.

	The slope must change somewhere below $1\,M_\odot$
because at low masses 
($M<0.6\,M_\odot$) and after correcting for binaries, several studies in
different environments all find a flat MF, $\alpha\sim 0$ (Paper IV;
Reid \& Gizis 1997; Scalo 1998; Holtzman et al.\ 1998), although different
authors argue for different break points.  In Paper IV, we found a break
at $M\sim 0.6\,M_\odot$ from {\it HST} counts of disk M dwarfs.  
Significantly, however,
this break point coincides with the boundary between our own
M dwarf data and the MF for higher mass stars derived by 
Wielen, Jahrei\ss, \& Kr\"uger (1983) from stars within 20 pc.  Reid \& Gizis
(1997) argue that their 8 pc sample is intrinsically cleaner than the 20 pc
sample of Wielen et al.\ (1983) and find that
the MF is flat all the way up to $1\,M_\odot$.  Scalo (1998) also adopts
$1\,M_\odot$ as the break point.  
However, Holtzman et al.\ (1998) find that the LF of Galactic
bulge stars in Baade's Window is very similar to the local disk LF derived
in Paper IV (including the higher-mass data from Wielen et al.\ 1983) and
thus also derive a similar MF.  For simplicity, we adopt a break at the
last point of our observations, that is $\alpha = -1.7$ for $M>0.71\,M_\odot$
and $\alpha=0.25\,M_\odot$ for $M<0.71\,M_\odot$.  In fact, our final results
do not depend strongly on the exact point of the transition.

\subsection{5.2.1 Mass Density Without Correction for Binaries}

	Taking equation \dndlnm\ at face value, the local mass density of 
the spheroid within the observed mass range $0.09< M/M_\odot < 0.71$ is
$$\rho_\obs = (2.86\pm 0.92)\times 10^{-5}\,M_\odot\,\pc^{-3}\qquad 
(0.09<M/M_\odot<0.71),\eqn\rhototone$$

	Extending equation \dndlnm\ into the brown dwarf regime, we find
that the total substellar density is 
$\rho_{bd} = 0.23\times 10^{-5}\,M_\odot$.
This is an order of magnitude
smaller than the stellar component of the spheroid evaluated in equation
\rhototone.  The statistical errors are more than 50\%.
However, the important
point is that substellar objects do not make a major contribution to the
spheroid mass density unless the slope of the mass function changes sharply
at the hydrogen-burning limit.

	Using equation \dndlnmhigh, we find the total mass of
stars in the range $0.71<M/M_\odot<0.9$ is
$\rho_{to} = 0.84\,\times 10^{-5}\,M_\odot\,\pc^{-3}$.  Thus, the mass
density of hydrogen-burning spheroid stars is $\rho_{hb}=
\rho_\obs + \rho_{to} = (3.7 \pm 1.0)\times 10^{-5}\,M_\odot\,\pc^{-3}$,
which can be compared to the value obtained by Bahcall, Schmidt, \& Soneira
(1983) of $\rho_{hb}= (4 - 14)\times 10^{-5}\,M_\odot\,\pc^{-3}$.  

Since the great
majority of remnants are white dwarfs, we adopt $M=0.6\,M_\odot$ for all of
the remnants of progenitors $M>0.9\,M_\odot$.  We find a remnant mass density
$\rho_{wd} = 1.3\times 10^{-5}\,M_\odot\,\pc^{-3}$, and hence a total mass
density
$$\rho_{\rm tot} = 5.2\times 10^{-5}\,M_\odot\,\pc^{-3}.\eqn\rhotottwo$$
The statistical errors associated with this estimate are about 50\%, but
the largest sources of uncertainty are the arbitrary assumptions used to
extend the mass function.  

We note
that had we chosen to break the power law at $1\,M_\odot$ rather than
at $0.71\,M_\odot$, the contribution of more massive stars and remnants would
have increased from
$\rho_{to}+\rho_{wd}=2.1\,\times 10^{-5}\,M_\odot\,\pc^{-3}$ to
$3.5\,\times 10^{-5}\,M_\odot\,\pc^{-3}$.  This would imply a 27\% increase
in the overall density.

\subsection{5.2.2 Correction for Binaries}

	We are not aware of any data on the binary fraction for spheroid M
dwarfs.  We therefore somewhat arbitrarily adopt a correction similar to the 
one we derived for disk M dwarfs (Paper IV).  For stars $(M<0.71\,M_\odot)$,
we decrease (steepen) the power law by $0.35$ to $\alpha=-0.10$, and
we fix the normalization at $0.71\,M_\odot$ to the uncorrected density.  This
yields
$${d N\over d\log M} = 12.5\times 10^{-5}
\biggl({M\over 0.71\,M_\odot}\biggr)^{-0.1}\,\pc^{-3}
\qquad (M<0.71\,M_\odot),\eqn\dndlnmcorr$$
For higher masses, we continue to use equation \dndlnmhigh.

We then find in units of $ 10^{-5}\,M_\odot\,\pc^{-3}$,  
$\rho_\obs = 3.6$,
$\rho_{bd} = 0.7$,
$\rho_{to} = 0.8$, and
$\rho_{wd} = 1.3$.  That is, the total density,
$$\rho_{\rm tot}= 6.4\times 10^{-5}\,M_\odot\,\pc^{-3}\qquad 
({\rm including}\ {\rm binaries}),\eqn\withbin$$
is only about 25\% higher than the uncorrected result.
For comparison, the ``heavy spheroid'' model of Caldwell \& Ostriker (1981)
predicts a local density of $111\times 10^{-5}\,M_\odot\,\pc^{-3}$.

Fuchs \& Jahrei\ss\ (1998) have obtained a lower limit to the local
mass density of spheroid subdwarfs of $1 \times 10^{-4}\,M_\odot\,\rm
pc^{-3}$ using reliable Hipparcos parallaxes of stars in the Fourth
Catalog of Nearby Stars (CNS4, Jahrei\ss\ \& Wielen 1997) for a somewhat
broader range of subdwarfs than we consider here. We can compare our
result of $\rho_{\rm obs} = 3.6 \times 10^{-5}$ M$_\odot$ pc$^{-3} $
with that of Fuchs \& Jahrei\ss\ (1998) as follows. We count stars from
their Table 1 in the range $0.09 < M/M_\odot < 0.71$ and in order to
guard against contamination by the intermediate population, we select
only stars with velocities $V_T> 220$ km\thinspace s$^{-1}$.  We then
divide by the completeness factor of 0.53 (see \S\ 4.3.2, but note that
the factor differs very slightly because of slightly different geometries
of the sample).  This
procedure yields 4 stars with total mass $M_{\rm tot}/0.53 = 2.1\, M_\odot$ 
within 25 pc or
$(3.2\pm 1.8) \times 10^{-5}$ M$_\odot$ pc$^{-3}$.  We note that there
are two additional spheroid stars lurking at 25.5 pc (B.\ Fuchs \&
H.\ Jahrei\ss\ 1998, private communication) just beyond the 25 pc distance
limit of the CNS4 catalog, which would raise the density to $(4\pm
2)\times 10^{-5}$ M$_\odot$ pc$^{-3}$.  These lower limits are consistent
within the errors with our estimate for the observed density
$\rho_{\rm obs} = 3.6 \times 10^{-5}$ M$_\odot$ pc$^{-3}$.

The local normalization of the dark halo is
$\rho_{\rm halo}\sim 9\times 10^{-3}\,M_\odot\,\pc^{-3}$.  Of order half
of this value may be in the form of compact objects now being detected in
microlensing observations toward the LMC (Alcock et al.\ 1997).  Thus, the
spheroid contributes only $\sim 1\%$ of the observed microlensing
optical depth.

If, as we discussed in \S\ 4.3.4, the spheroid is composed of two components, 
one highly flattened and one roughly spherical, then our results would
be sensitive only to the latter.  In this case, the local density would be
higher by a factor $\sim 5/3$.  This higher density would not affect the
spheroid's microlensing optical depth, however, because the flattened component
would not contribute significantly to microlensing.

{\bf Acknowledgements}:  This paper benefits greatly from conversations
and correspondence sparked by its appearance as a preprint.  Conard Dahn
and Neill Reid (before he became referee) independently pointed out an
important computational error in the renormalization of the DLHG LF.  Ivan
King drew our attention to the work of Capaccioli et al.  Isabelle Baraffe
and Gilles Chabrier helped clarify several issues related to theoretical
isochrones.  We are grateful to Conard Dahn and Isabelle Baraffe for making
available some of their work in advance of publication.  Finally, we are
indebted to Neill Reid for a very thorough and helpful referee report.
A.\ G.\ was supported in part
by NSF grant AST 9420746 and in part by NASA grant NAG5-3111.
J.\ N.\ B.\ was supported by NASA grant NAG5-1618.
The work is based in large part
on observations with the NASA/ESA Hubble Space Telescope, obtained
at the Space Telescope Science Institute, which is operated by the
Association of Universities for Research in Astronomy, Inc. (AURA), under
NASA contract NAS5-26555.  Important supplementary observations were made
at KPNO and CTIO operated by AURA.

\endpage

\Ref\Alcock{Alcock, C., et al.\ 1997, ApJ, 486, 697}
\Ref\Bst{Bahcall, J.\ N.\ 1986, ARA\&A, 24, 577}
\Ref\BC{Bahcall, J.\ N., \& Casertano, S.\ 1986, ApJ, 308, 347}
\Ref\BFG{Bahcall, J.\ N., Flynn, C., Gould, A., \& Kirhakos, S.\ 1994, ApJ, 
435, L51 (Paper I)}
\Ref\BKSS{Bahcall, J.\ N., Kirhakos, S., Saxe, D.\ H., \& Schneider, D.\ P.\
1997, ApJ, 479, 642}
\Ref\BSS{Bahcall, J.\ N., Schmidt, M., \& Soneira, R.\ M.\ 1983, ApJ, 265, 730}
\Ref\BS{Bahcall, J.\ N.\ \& Soneira, R.\ M.\ 1980, ApJS, 44, 73}
\Ref\BC{Baraffe, I., \& Chabrier, G.\ 1996, ApJ, 461, L51}
\Ref\BC{Baraffe, I., Chabrier, G.\ Allard, F., \& Hauschildt, P.\ H.\ 1995, 
ApJ, 446, L35}
\Ref\BC{Baraffe, I., Chabrier, G.\ Allard, F., \& Hauschildt P.\ H.\ 1997,
A\&A, 327, 1057}
\Ref\BSL{Beers, T. \& Sommer-Larsen, J., 1995, ApJS, 96, 175}
\Ref\co{Caldwell, J.\ A.\ R.\ \& Ostriker, J.\ P.\ 1981, ApJ, 251, 61}
\Ref\cps{Capaccioli, M., Piotto, G., \& Stiavelli, M.\ 1993, MNRAS, 261, 819}
\Ref\crb{Casertano, S., Ratnatunga, K.\ U., \& Bahcall, J.\ N.\ 1990, ApJ,
357, 435}
\Ref\cm{Chabrier, G., \& M\'era, D.\ 1997, A\&A, 328, 83}
\Ref\chy{Chiba, M.\ \& Yoshii, Y.\ 1998, AJ, 115, 229}
\Ref\dah{Dahn, C.~C., Liebert, J.\ W., Harris, H., \& Guetter, H.\ C.\ 1995, 
p.\ 239, An ESO Workshop on: the Bottom of the Main Sequence and Beyond,
C. G. Tinney ed. (Heidelberg: Springer)}
\Ref\dm{D'Antona, F., \& Mazzitelli, I.\ 1996, ApJ, 456, 329}
\Ref\egg{Eggen, O.\ J.\ 1979a, ApJS, 39, 89}
\Ref\egg{Eggen, O.\ J.\ 1979b, ApJ, 230, 786}
\Ref\egg{Eggen, O.\ J.\ 1983, ApJS, 43, 457}
\Ref\ff{Flynn, C.\ \& Gould, A., \& Bahcall, J.\ N., 1996, ApJ, 466, L55 
(Paper II)}
\Ref\fj{Fuchs, B.\ \& Jahrei\ss, H.\ 1998, A\&A, 329, 81}
\Ref\gil{Gilmore, G.\ 1989, ARA\&A, 27, 555}
\Ref\gizis{Gizis, J.\ E.\ 1997, AJ, 113, 806}
\Ref\gbm{Gould, A., Bahcall, J.\ N., \& Flynn, C.\ 1996, ApJ, 465, 759 
(Paper III)}
\Ref\gbm{Gould, A., Bahcall, J.\ N., \& Flynn, C.\ 1997, ApJ, 482, 913 
(Paper IV)}
\Ref\gf{Graff, D.\ S.\ \& Freese, K.\ 1996, ApJ, 456, L49}
\Ref\harris{Harris, W.\ E.\ 1976, AJ, 81, 1095}
\Ref\hart{Hartwick, R.\ D.\ A.\ 1987, in The Galaxy, G.\ Gilmore and B.\
Carswell, eds., p.\ 281 (Dordrecht: Reidel)}
\Ref\hart{Hartwick, R.\ D.\ A.\ \& Schade, D.\ 1990, ARA\&A, 28, 437}
\Ref\hm{Henry, T.\ J., \& McCarthy, D.\ W.\ Jr.\ AJ, 106, 773}
\Ref\holtz{Holtzman, J.\ A., Watson, A.\ M., Baum, W.\ A., Grillmair, C.\ J.,
Groth, E.\ J., Light, R.\ M., Lynds, R., \& O'Neil, E.\ J.\ 1998, AJ, in press
(astro-ph9801321)}
\Ref\jw{Jahrei\ss, H. \& Wielen, R., 1997, in: Presentation
of the HIPPARCOS and TYCHO catalogues, eds. M.A.C.
Perryman, P.L. Bernacca, ESA-SP 402, (Nordwijk:ESTEC)}

\Ref\jones{Jones, J.\ B., Driver, S.\ P.,
  Phillipps, S, Davies, J.\ I., Morgan, I., \& Disney, M.\ J.\ 1997, A\&A,
318, 729}
\Ref\kinman{Kinman, T.\ D., Wirtanen, C.\ A., \& Janes, K.\ A.\ 1965, ApJS,
11, 223}
\Ref\ktg{Kroupa, P., Tout, C.\ A., \& Gilmore, G.\ 1993, MNRAS, 262, 545}
\Ref\laird{Laird, J.\ B., Carney, B.\ W., Rupen, M.\ P., \& Latham, D.\ W.\ 
1988, 96, 108}
\Ref\lay{Layden, A.\ C., Hanson, R.\ B., Hawley, S.\ L., Klemola, A.\ R.,
Hanley, C.\ J.\ 1996, AJ, 112 2110}
\Ref\ldm{Liebert, J., Dahn, C.\ C., \& Monet, D.\ G.\ 1988, ApJ, 332, 891}
\Ref\lowen{Lowenthal, J.\ D., Koo, D.\ C., Guzman, R., Gallego, J.,
Phillips, A.\ C., Faber, S.\ M., Vogt, N.\ P., Illingworth, G.\ D., \&
Gronwall, C.\ 1997, ApJ, 481, 673}
\Ref\mor{Morrison, H.\ 1993, AJ, 106, 587}
\Ref\mendez{M\'endez, R.\ A., Minniti, D., De Marchi, G., Baker, A., \&
Couch, W.\ J.\ 1997, MNRAS 283, 666}
\Ref\mcs{M\'era, D., Chabrier, G., \& Schaeffer, R.\ 1996, Europhys.\ Lett.,
33, 327}
\Ref\ns{Nissen, P.\ E.\ \& Schuster, W.\ J.\ 1991, A\&A, 251, 457}
\Ref\pck{Piotto, G., Cool, A.\ M., \& King, I.\ R.\ 1997, AJ, 113, 1345}
\Ref\pow{Popowski, P.\ \& Gould, A.\ 1998, ApJ, submitted (astro-ph 9802168)}
\Ref\psb{Preston, G.\ W., Schectman, S.\ A., \& Beers, T.\ C.\ 1991, ApJ,
375,121}
\Ref\rg{Reid, I.\ N.\ \& Gizis, J.\ E.\ 1997, AJ, 113, 2246}
\Ref\rei{Reid, I.\ N.,  Hawley, S.\ L., \& Gizis, J.\ E.\ 1995, AJ, 110, 1838}
\Ref\rymts{Reid, I.\ N., Yan, L., Majewski, S., Thompson, I., \& Smail, I.\
1996, AJ, 112, 1472}
\Ref\reid{Reid, M., J.\ 1993, ARA\&A, 31, 345}
\Ref\RF{Richer, H.\ B.\ \& Fahlman, G.\ G.\ 1992, Nature, 358, 383}
\Ref\scalo{Scalo, J.\ A.\ 1998, in The Stellar Initial Mass Function 
Proceedings of the 38th Herstmonceux Conference, eds.\ G.\ Gilmore, I.\ 
Parry, and S.\ Ryan, in press}
\Ref\schmidt{Schmidt, M.\ 1975, 202, 22}
\Ref\slz{Sommer-Larsen, J.\ \& Zhen, C.\ 1990, MNRAS, 242, 10}
\Ref\cps{Stiavelli, M., Piotto, G., Capaccioli, M., \& Ortolani, S.\ 1991, 
in Janes K., ed., Formation and Evolution of Star Clusters, p.\ 449 
(Boston, ASP)}
\Ref\steidel{Steidel, C., Giavalisco, M., Dickenson, M., \& Adelberger, K.\
L.\ 1996, AJ, 112, 352}
\Ref\wieetal{Stobie, R.\ S., Ishida, K., \& Peacock, J.\ A.\ 1989, MNRAS, 
238, 709}
\Ref\WJK{Wielen, R., Jahrei\ss, H., \& Kr\"uger, R.\ 1983, IAU Coll.\ 76:
Nearby Stars and
the Stellar Luminosity Function, A.\ G.\ D.\ Philip and A.\ R.\ Upgren eds.,
p 163}
\Ref\zinn{Zinn, R.\ 1985, ApJ, 293, 424}
\refout
\endpage
%\figout
\end